# Experimental Observation of Surface States and Landau Levels Bending in Bilayer Graphene

Long-Jing Yin[§], Yu Zhang[§], Jia-Bin Qiao, Si-Yu Li, and Lin He*

Center for Advanced Quantum Studies, Department of Physics, Beijing Normal University, Beijing, 100875, People's Republic of China

[§]These authors contributed equally to this work.
* Email: helin@bnu.edu.cn

**We report on microscopic measurements of the low-energy electronic structures both at zigzag and armchair edges of bilayer graphene using scanning tunneling microscopy and spectroscopy (STM and STS). We have found that, both in the absence and in the presence of a magnetic field, an almost zero-energy peak in density of states was localized at zigzag edges, as expected for the surface states at zigzag edges of bilayer graphene. In the quantum Hall regime, we have observed clearly Landau levels bending away from the charge neutrality point near both the zigzag and armchair edges. Such a result is a direct evidence for the evolution of Landau levels into the quantum Hall edge states in graphene bilayers. Our experiment indicates that it is possible to explore rich quantum Hall physics in graphene systems using STM and STS.**

There are two possible (perfect) edge terminations, i.e., zigzag and armchair, in graphene monolayer and the edge orientations affect the electronic structures of graphene sheet strongly [1,2]. Very recently, graphene system with zigzag edge termination has attracted much attention because that its surface states are believed to be closely related to the magnetic order and exceptional ballistic transport [3-5]. In the quantum Hall regime, both the zigzag and armchair edges could bend Landau levels (LLs) to produce dispersive edge states [6-8], which carry the chiral Dirac fermions responsible for the quantum Hall effect in graphene monolayer [9,10].

In graphene bilayers, the zigzag edge is also predicted to host surface states, but with an enhanced penetration into the bulk comparing to that of graphene monolayer [11]. In the presence of high magnetic fields, unconventional quantum Hall effect and a wealth of exotic electronic behavior have been observed in graphene bilayers [12-18]. Though many great successes have been achieved in the study of electronic properties of graphene bilayer, considerable work is still necessary to address some fundamental problems in this system. For example, a direct experimental observation of the surface states at the zigzag edges of graphene bilayer and its LLs bending at the edge terminations is still lacking up to now. In this Letter, we present scanning tunneling microscopy (STM) and spectroscopy (STS) measurements of bilayer graphene on graphite substrate both in the absence and in the presence of magnetic fields. The high-quality bilayer sample and the ultra-low random potential fluctuations due to substrate imperfections ensure us to direct probe the surface states at the zigzag edges and to measure the LLs bending at both the zigzag and armchair edges.

We performed STM measurements in an UNISOKU (USM-1500S) instrument with the magnetic fields up to 8 T. The STS spectra, *i.e.*, differential conductance $dI/dV$ curves, were measured with a lock-in detection (modulation voltage: 5-10 mV, frequency: 793 Hz). The STM tips were obtained by chemical etching from a wire of $Pt_{0.8}Ir_{0.2}$ alloys. Lateral dimensions observed in the STM images were calibrated using a standard graphene lattice and Ag (111) surface. All the STM and STS measurements were performed in the ultrahigh vacuum chamber (~ $10^{-11}$ Torr) at ~ 4.4 K. Bilayer graphene samples used in our experiments were prepared on ZYA grade (NT-MDT) Highly Oriented Pyrolytic Graphite (HOPG) substrates. The HOPG samples were surface cleaved by adhesive tape prior to experiments immediately. The bilayer graphene flakes deposited on the substrate during the process of mechanical exfoliation and, very importantly, these graphene sheets may decouple from the graphite surface, as demonstrated in previous studies [19-25].

Figure 1(a) shows a representative STM image of a Bernal graphene bilayer on graphite surface. The triangular contrast in the atomic image arises from the A/B atoms' asymmetry generated by the adjacent two layers. To further identify the bilayer graphene region, we carried out STS measurement in various magnetic fields, as shown in Fig. 1(b). In zero magnetic field, there are two peaks located at about 0 mV and 25 mV in the spectrum, which are attributed to the density of states (DOS) peak generated at the valence-band edge (VBE) and conduction-band edge (CBE) of a gapped bilayer respectively [22,26]. The finite gap in the low-energy bands is generated by inversion symmetry breaking of the adjacent two layers induced by the substrate [22,26-33].

The spectra recorded in high magnetic fields [Fig. 1(b)] exhibit Landau quantization of massive Dirac fermions, as expected for the gapped graphene bilayers [13,22,26]. The LL sequences of gapped graphene bilayers can be described by

$$E_n = E_C \pm [(\hbar\omega_c)^2(n(n-1)) + (U/2)]^{1/2} - \xi z U/4, \qquad n = 2,3,4\ldots$$

$$E_0 = E_C + \xi U/2, \qquad E_1 = E_C + \xi (U/2)(1-z), \tag{1}$$

where $E_C$ is the energy of charge neutrality point (CNP), $\omega_c = eB/m^*$ is the cyclotron frequency, $m^*$ is the effective mass of charge carriers, and $\xi = \pm$ are the valley indices. We have $z = 2\hbar\omega_c/t_\perp \ll 1$ for $B \leq 8$ T and $|U| \approx E_g$ (gap energy) when the interlayer bias $U < t_\perp$. According to the fitting, as shown in Fig. 1(c), we obtain $E_g \approx 25$ meV and $m^* = (0.035 \pm 0.002)m_e$ ($m_e$ is the free-electron mass). Both the values of $E_g$ and $m^*$ agree well with the range of values reported in Bernal bilayers previously [22,26]. Note that the two lowest levels $LL_{(0,1,+)}$ and $LL_{(0,1,-)}$ are a couple of layer-polarized quartets, and they are mainly localized on the first and second graphene layers, respectively. Therefore, the signal of $LL_{(0,1,+)}$ is much stronger than that of $LL_{(0,1,-)}$ in the spectra since that the STS predominantly probes the DOS on the top layer.

The said measurements demonstrate explicitly that the topmost two layers are Bernal stacking and they are completely decoupled from the substrate. Once a high quality bilayer graphene region is identified, the structures and electronic properties around its edges are carefully studied. Fig. 2(a) and 2(b) show typical atomic-resolution images of a zigzag edge and an armchair edge of the graphene bilayer, respectively. Away from the edges, the STM images exhibit triangular contrasting, as expected to be observed in Bernal bilayer. The types of the terminative edges can be determined by the arrangement of the triangular dots, as schematically shown in Fig. 2(a-c). Around both the zigzag and armchair edges, clear interference patterns are observed. Similar interference patterns have also been observed around edges of graphene monolayer on graphite substrate [34-36] and are attributed to the interference between the incident and scattered electron waves in two Dirac cones at the atomically sharped boundaries (see Fig. S1 in Supplementary Information [37] for more experimental data). The graphene bilayers with zigzag and armchair edges are expected to exhibit quite different electronic band structures: there are localized surface states at zigzag edges but not at armchair edges [11,38], as shown in Fig. 2(d)-2(f). In a gapped graphene bilayer with zigzag edges, the surface states may be layer-polarized [38] and they are predicted to have a much larger penetration length into the bulk than that in graphene monolayer [11].

To study the effect of edges on the electronic properties of bilayer graphene, we measured the spatial-resolved $dI/dV$ spectra near both the zigzag and armchair edges under zero magnetic field. Fig. 3 shows a representative result obtained around a zigzag edge (see Fig. S2 in Supplementary Information [37] for experimental data recorded around an armchair edge). Typical tunneling spectra recorded at different distances away from the edge are shown in Fig. 3(a) [here we define zero-position at the zigzag edge, as shown in Fig. 3(b)]. With approaching the edge, the signal of the DOS peaks at the VBE and CBE [dashed lines in Fig. 3(a)] becomes weak because that the VBE and CBE of a gapped bilayer graphene are exactly valid in the bulk and they are less well defined around the edges. The energy spacing between the VBE and the CBE increases about 10 meV [Fig. 3(a)], which may arise from a slightly enhanced band gap around the edge. Beside the said result, another notable feature of the spectra is the emergence of a new DOS peak around the zigzag edge and the signal of the peak increases with approaching the edge. Such a peak, which is absent around the armchair edge (Figure S2), is attributed to a layer-polarized surface state at the zigzag edge of the gapped graphene bilayer [38]. The presence of this DOS peak (surface state) is a fundamental result, although anticipated in many theoretical works, had never been experimentally observed before in graphene bilayer. Fig. 3(c) plots the measured peak height of the surface state as a function of the distance from the edge. The surface state shows a decreasing intensity with increasing distance and extends over 10 nm away from the edge, consistent with the expected decaying behavior of the surface states in graphene bilayer. Here we should point out that the decaying length of the surface states in graphene bilayer is much larger than that in graphene monolayer [Fig. 3(c)]. Additionally, the surface state can also be detected even in the quantum Hall regime [Fig. 4(a) and 4(b)].

In two-dimensional electron systems, the low-energy band structures of quasi-particles develop into dispersionless LLs in the presence of a high magnetic field and give raise to the insulating behavior in the bulk. While the confining potential at the edges of the system bends the discrete LLs to form dispersive edge states that carry charge carries in the quantum Hall effect. The high-quality bilayer sample with crystallographically perfect edges and the ultra-low random potential fluctuations induced by the substrate, as demonstrated in Fig. 1-3, allow us to direct probe the LLs bending at the edges.

Figure 4 summarizes the measured result of bilayer graphene in the quantum Hall regime and we observe clearly LLs bending at both the zigzag and armchair edges (see Fig. S3 in Supplementary Information [37] for more experimental data). Away from the edges, the well-defined LL spectra, as shown in Fig. 4(a)-4(d), follow the sequence of massive Dirac fermions in gapped graphene bilayers (here we use $l_B = \sqrt{\hbar/eB}$ to define the distance from the edge). With approaching the edges, the DOS peaks for the LLs become weak and the LLs are shifted away from the charge neutrality point, as shown in Fig. 4(e) and 4(f). At a fixed energy, the measured local DOS at position *r* is determined by the wavefunctions according to $\rho(r) \propto |\psi(r)|^2$, while the wavefunctions of LLs have their spatial extent, $\sim 2\sqrt{N} l_B$. It indicates that there is an important contribution from the bulk states even for the recorded LL spectra near the edges. The wavefunctions of LLs with higher indices have greater spatial extents, as shown in the inset of Fig. 4(f). Consequently, the amplitude of high-index LL peaks decreases slower than that of low-index LL peaks [Fig. 4(e)] and the bending of low-index LLs seems stronger than that of high-index LLs [Fig. 4(f)] (due to greater contribution from the bulk states to higher LLs).

Theoretically, the shift length of the LLs bending around the edges is predicted to be of the magnetic length [7,8]. In Fig. 4(g), we summarized the measured shift length at different magnetic fields around both the zigzag and armchair edges. We find that the shift length depends on neither the magnetic fields nor the edge types, and it is of the magnetic length (see Fig. S4 in Supplementary Information [37] for more experimental data). Additionally, the shift length seems to be dependent on the LL index: the estimated shift lengths for the $LL_{(0,1,+)}$ and $LL_2$ are about $\sim 1.4\ l_B$ and $\sim 2.0\ l_B$, respectively.

In conclusion, we measured the surface state and its spatial evolution around the zigzag edges of bilayer graphene. Our result demonstrated an enhanced penetration length of the surface states in bilayer graphene comparing to that in graphene monolayer. In the quantum Hall regime, we provided direct evidence for the LLs bending around both the zigzag and armchair edges of bilayer graphene, which may open the door to explore exotic quantum Hall physics in graphene bilayers using scanned probe techniques.

**Acknowledgments**

This work was supported by the National Basic Research Program of China (Grants Nos. 2014CB920903, 2013CBA01603), the National Natural Science Foundation of China (Grant Nos. 11422430, 11374035), the program for New Century Excellent Talents in University of the Ministry of Education of China (Grant No. NCET-13-0054), Beijing Higher Education Young Elite Teacher Project (Grant No. YETP0238). L.H. also acknowledges support from the National Program for Support of Top-notch Young Professionals.

**Reference:**

[1] A. H. Castro Neto, F. Guinea, N. M. R. Peres, K. S. Novoselov, A. K. Geim, The electronic properties of graphene. *Rev. Mod. Phys.* **81**, 109-162 (2009).

[2] M. O. Goerbig, Electronic properties of graphene in a strong magnetic field. *Rev. Mod. Phys.* **83**, 1193-1243 (2011).

[3] Y.-W. Son, M. L. Cohen, S. G. Louie, Half-metallic graphene nanoribbons. *Nature* **444**, 347 (2006).

[4] G. Z. Magda, X. Jin, I. Hagymasi, P. Vancso, Z. Osvath, P. Nemes-Incze, C. Hwang, L. P. Biro, L. Tapaszto, Room-temperature magnetic order on zigzag edges of narrow graphene nanoribbons. *Nature* **514**, 608 (2014).

[5] J. Baringhaus, M. Ruan, F. Edler, A. Tejeda, M. Sicot, A. Taleb-Ibrahimi, A.-P. Li, Z. Jiang, E. H. Conrad, C. Berger, C. Tegenkamp, W. A. de Heer, Exceptional ballistic transport in epitaxial graphene nanoribbons. *Nature* **506**, 349 (2014).

[6] V. P. Gusynin, S. G. Sharapov, Unconventional integer quantum Hall effect in graphene. *Phys. Rev. Lett.* **95**, 146801 (2005).

[7] D. A. Abanin, P. A. Lee, L. S. Levitov, Spin-filtered edge states and quantum Hall effect in graphene. *Phys. Rev. Lett.* **96**, 176803 (2006).

[8] G. Li, A. Luican-Mayer, D. Abanin, L. Levitov, and E. Y. Andrei, Evolution of Landau levels into edge states in graphene. *Nature Commun.* **4**, 1744 (2013).

[9] K. S. Novoselov, A. K. Geim, S. V. Morozov, D. Jiang, M. I. Katsnelson, I. V. Grigorieva, S. V. Dubonos, A. A. Firsov, Two-dimensional gas of massless Dirac fermions in graphene. *Nature* **438**, 197 (2005).

[10] Y. Zhang, Y.-W. Tan, H. L. Stormer, P. Kim, Experimental observation of quantum Hall effect and Berry's phase in graphene. *Nature* **438**, 201 (2005).

[11] E. V. Castro, N. M. R. Peres, J. M. B. Lopes dos Santos, A. H. Castro Neto, and F. Guinea, Localized states at zigzag edges of bilayer graphene. *Phys. Rev. Lett.* **100**, 026802 (2008).

[12] K. S. Novoselov, E. McCann, S. V. Morozov, V. I. Fal'ko, M. I. Katsnelson, U. Zeitler, D. Jiang, F. Schedin, A. K. Geim, Unconventional quantum Hall effect and Berry's phase of $2\pi$ in bilayer graphene. *Nature Phys.* **2**, 177 (2006).

[13] E. McCann, V. Fal'ko, Landau-Level Degeneracy and Quantum Hall Effect in a Graphite Bilayer. *Phys. Rev. Lett.* **96**, 086805 (2006).

[14] F. Zhang, A. H. MacDonald, Distinguishing Spontaneous Quantum Hall States in Bilayer Graphene. *Phys. Rev. Lett.* **108**, 186804 (2012).

[15] B. J. LeRoy, M. Yankowitz, Emergent complex states in bilayer graphene. *Science* **345**, 31-32 (2014).

[16] A. Kou, *et al.* Electron-hole asymmetric integer and fractional quantum Hall effect in bilayer graphene. *Science* **345**, 55-57 (2014).

[17] K. Lee, *et al.* Chemical potential and quantum Hall ferromagnetism in bilayer graphene. *Science* **345**, 58-61 (2014).

[18] P. Maher, *et al.* Tunable fractional quantum Hall phases in bilayer graphene. *Science* **345**, 61-64 (2014).

[19] G. Li, A. Luican, E. Y. Andrei, Scanning Tunneling Spectroscopy of Graphene on Graphite. *Phys. Rev. Lett.* **102**, 176804 (2009).

[20] G. Li, E. Y. Andrei, Observation of Landau levels of Dirac fermions in graphite. *Nat. Phys.* **3**, 623-627 (2007).

[21] L.-J. Yin, *et al.* Tuning structures and electronic spectra of graphene layers with tilt grain boundaries. *Phys. Rev. B* **89**, 205410 (2014).

[22] L.-J. Yin, S.-Y. Li, J.-B. Qiao, J.-C. Nie, and L. He, Landau quantization in graphene monolayer, Bernal bilayer, and Bernal trilayer on graphite surface. *Phys. Rev. B* **91**, 115405 (2015).

[23] L.-J. Yin, J.-B. Qiao, W.-J. Zuo, W.-T. Li, and L. He, Experimental evidence for non-Abelian gauge potentials in twisted graphene bilayers. *Phys. Rev. B* **92**, 081406(R) (2015).

[24] G. Li, A. Luican-Mayer, D. Abanin, L. Levitov, E. Y. Andrei, Evolution of Landau levels into edge states in graphene. *Nature Commun.* **4**, 1744 (2013).

[25] W.-X. Wang, *et al.* Atomic resolution imaging of the two-components Dirac Landau levels in graphene monolayer. *Phys. Rev. B* **92**, 165420 (2015).

[26] G. M. Rutter, *et al.* Microscopic polarization in bilayer graphene. *Nat. Phys.* **7**, 649-655 (2011).

[27] E. McCann, Asymmetry gap in the electronic band structure of bilayer graphene. *Phys. Rev. B* **74**, 161403(R) (2006).

[28] Y. Zhang, *et al.* Direct observation of a widely tunable bandgap in bilayer graphene. *Nature* **459**, 20-823 (2009).

[29] E. Castro, *et al.* Biased Bilayer Graphene: Semiconductor with a Gap Tunable by the Electric Field Effect. *Phys. Rev. Lett.* **99**, 216802 (2007).

[30] T. Ohta, A. Bostwick, T. Seyller, K. Horn, E. Rotenberg, Controlling the electronic structure of bilayer graphene. *Science* **313**, 951-954 (2006).

[31] K. F. Mak, C. H. Lui, J. Shan, T. F. Heinz, Observation of an Electric-Field-Induced Band Gap in Bilayer Graphene by Infrared Spectroscopy. *Phys. Rev. Lett.* **102**, 256405 (2009).

[32] Z. Q. Li, *et al.* Band Structure Asymmetry of Bilayer Graphene Revealed by Infrared Spectroscopy. *Phys. Rev. Lett.* **102**, 037403 (2009).

[33] J. B. Oostinga, H. B. Heersche, X. Liu, A. F. Morpurgo, L. M. Vandersypen, Gate-induced insulating state in bilayer graphene devices. *Nature Mater* **7**, 151-157 (2008).

[34] Y. Niimi, T. Matsui, H. Kambara, K. Tagami, M. Tsukada, and H. Fukuyama, Scanning tunneling microscopy and spectroscopy of the electronic local density of states of graphite surfaces near monoatomic step edges. *Phys. Rev. B* **73**, 085421 (2006).

[35] Y. Kobayashi, K.-i. Fukui, and T. Enoki, Edge state on hydrogen-terminated graphite edges investigated by scanning tunneling microscopy. *Phys. Rev. B* **73**, 125415 (2006).

[36] Y. Kobayashi, K.-i. Fukui, T. Enoki, K. Kusakabe, and Y. Kaburagi, Observation of zigzag and armchair edges of graphite using scanning tunneling microscopy and spectroscopy. *Phys. Rev. B* **71**, 193406 (2005).

[37] See Supplemental Material for more STM images, STS spectra, and the detail of the analysis.

[38] W. Yao, S. Yang, and Q. Niu, Edge states in graphene: from gapped flat-band to gapless chiral modes. *Phys. Rev. Lett.* **102**, 096801 (2009).

**Figure captions:**

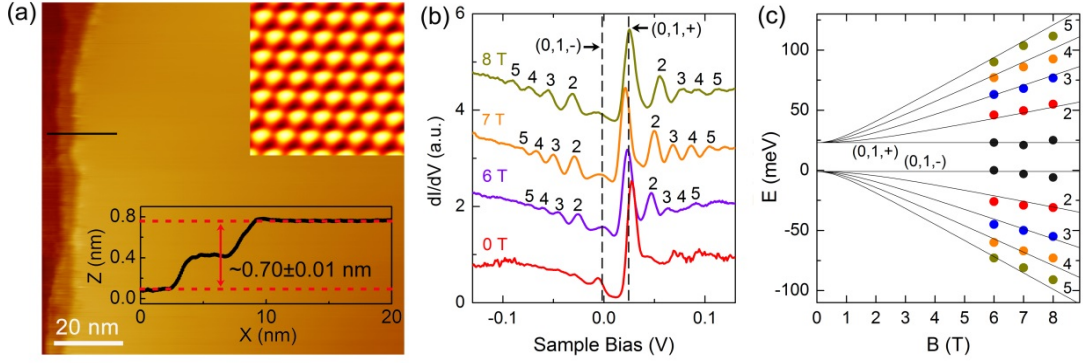

**FIG. 1** (color online). (a) 100 nm × 100 nm STM topographic image of a bilayer graphene region on graphite surface ($V_b$ = 0.2 V, $I$ = 0.2 nA). Inset: (upper) Atomic resolution image of the graphene bilayer showing the triangular contrasting, which reflects only one of the two sublattices of the topmost graphene due to the inversion symmetry breaking in Bernal (AB-stacked) bilayers; (lower) Height profile along the black line shows the height difference of two steps ~ (0.70 ± 0.01) nm, which is slightly larger than the equilibrium spacing of the bilayer step (~ 0.67 nm). (b) Tunneling spectra of the graphene bilayers recorded away from the edges under various magnetic fields. LL peak indices are labeled (+/- are valley indices) and the data are offset in Y-axis for clarity. (c) The LL peaks energies extracted from (b) plotted versus the magnetic fields $B$. The solid curves are the fitting of the data with Eq. (1).

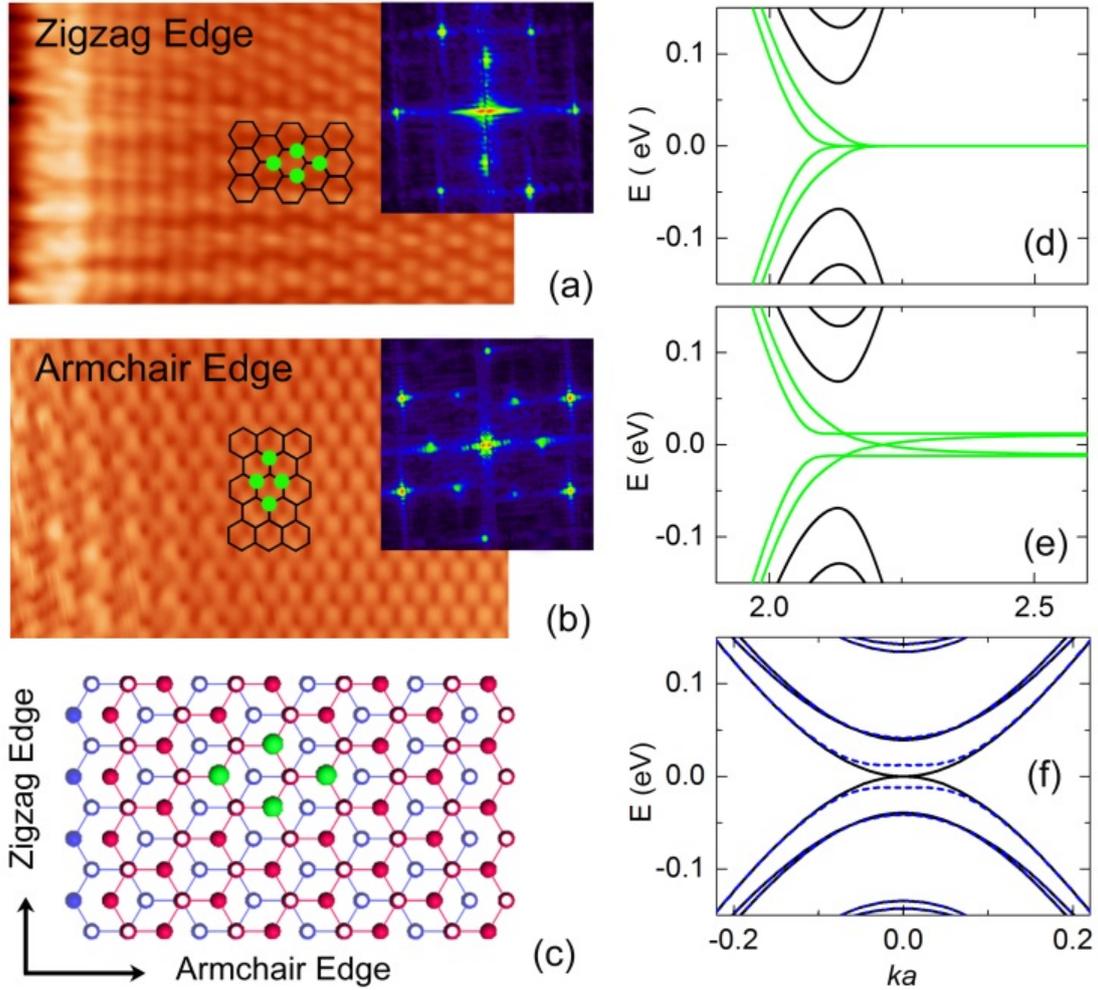

**FIG. 2** (color online). Atomic resolution images of a zigzag bilayer edge (a) and an armchair bilayer edge (b). The insets are the the fast Fourier transforms (FFT) of the STM images. The outer hexangular spots and inner bright spots correspond to the reciprocal lattice of the graphene lattice and the interference of the scattering, respectively. (c) Schematic of the Bernal bilayer graphene with the zigzag and armchair edges. The green dots, representing a set of sublattices imaged in STM topography, can be used to determine the type of graphene edge. Energy spectrum of an unbiased (d) and biased (e) bilayer graphene with zigzag edges. The green curves correspond to the quasi-localized surface states of the two zigzag edges. (f) Energy spectrum of an unbiased (solid lines) and biased (dotted lines) bilayer graphene ribbon with armchair edges.

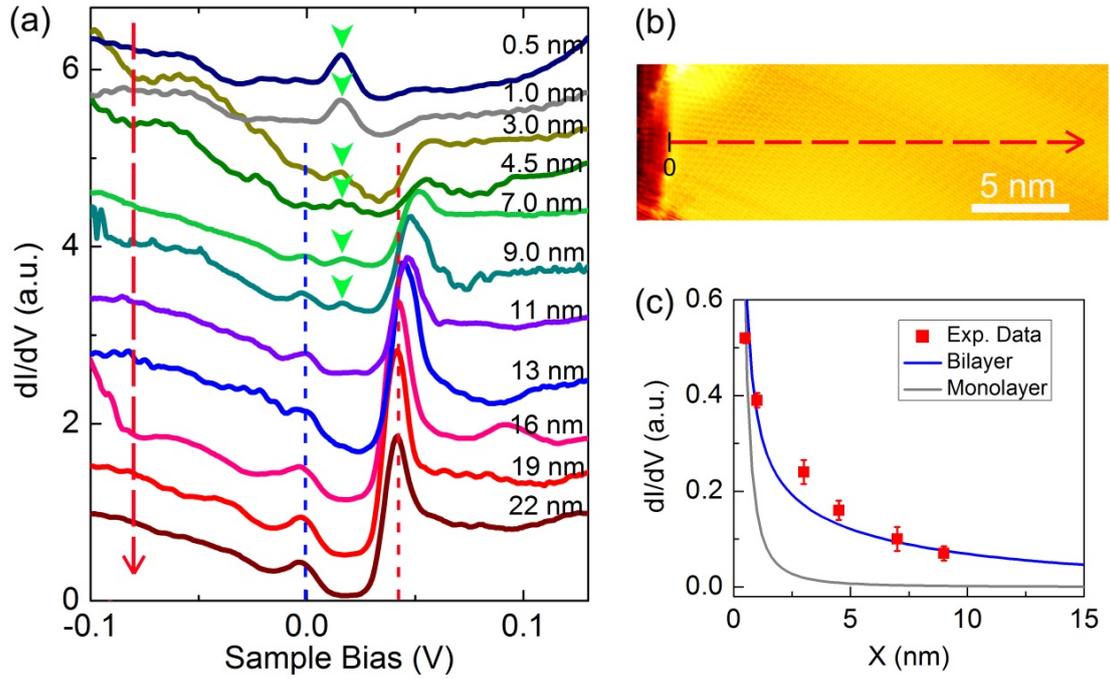

**FIG. 3** (color online). (a) Spatial-resolved STS spectra recorded along the line drawn in panel (b) around the zigzag edge of bilayer graphene under 0 T. The DOS peaks marked by green arrows correspond to the quasi-localized surface states of the zigzag edge in graphene bilayer. The blue and red dashed lines label the positions of the VBE and the CBE in the bulk graphene bilayer. (c) The decay of the DOS peak height of the surface state obtained in panel (a). The green and gray curves correspond to the expected decaying behavior of the surface states in graphene bilayer and monolayer, respectively.

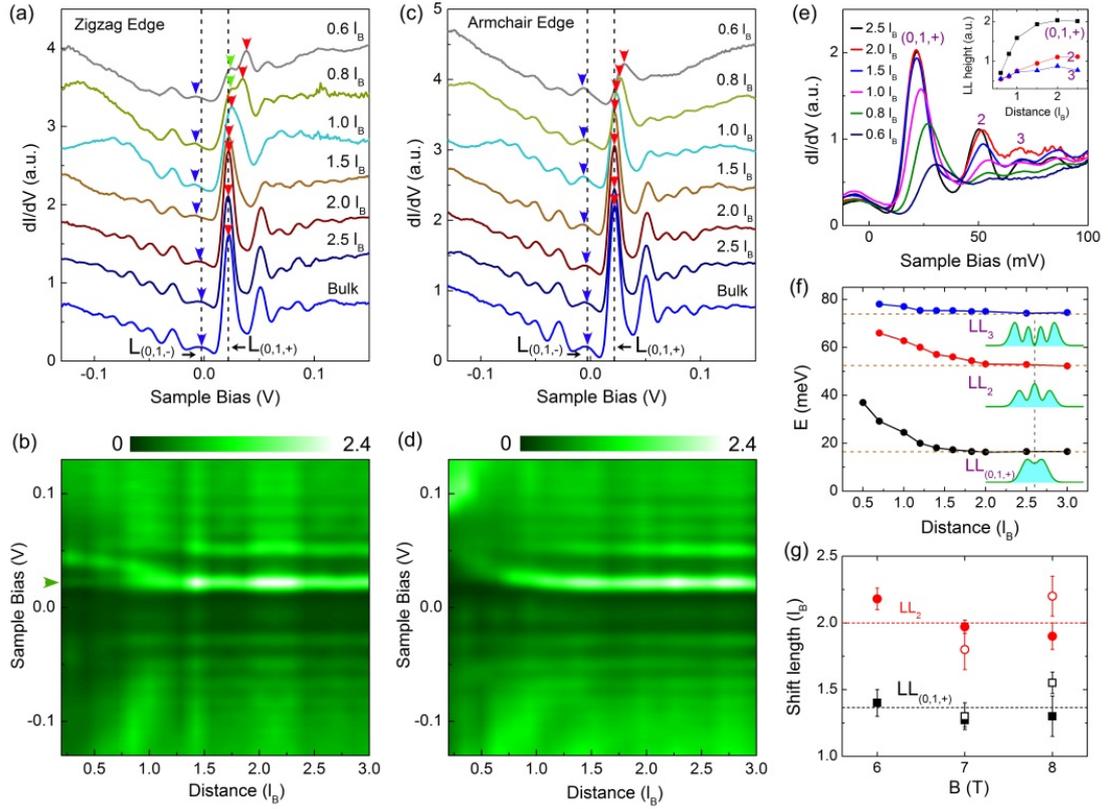

**FIG. 4** (color online). (a) and (c) show spatial variation of the LL spectra measured at 7 T around the zigzag and armchair edges of bilayer graphene, respectively. The dashed lines indicate the energy positions of the $LL_{(0,1,-)}$ and $LL_{(0,1,+)}$ in the bulk of bilayer graphene. The blue and red arrows mark the spatially evolution of the $LL_{(0,1,-)}$ and $LL_{(0,1,+)}$ peaks. (b) and (d) show LL spectra maps at 7 T recorded around the zigzag and armchair edges, respectively. In panel (a) and (b), the peaks marked by green arrows correspond to the quasi-localized surface states of the zigzag edge. (e) Evolution of the peak positions and heights at 7 T with distance from the armchair edge on the conduction-band side. Inset shows LL peak heights extracted from (e) as a function of the distance from the edge. (f) LL bending as a function of distance around the armchair edge measured at 8 T. It shows an explicit shift of the energy positions for the $LL_{(0,1,+)}$, $LL_2$ and $LL_3$ toward high energy with approaching the edge. The insets show calculated probability densities for the wave functions of the $LL_{(0,1,+)}$, $LL_2$ and $LL_3$ at 8 T. (g) Shift length of the $LL_{(0,1,+)}$ and $LL_2$ bending from the bilayer edges taken at different magnetic fields. The solid dots (empty dots) correspond to the data of armchair edges (zigzag edges). The dashed lines are the average values of ~

1.4 $l_B$ and ~ 2.0 $l_B$ for the $LL_{(0,1,+)}$ and $LL_2$, respectively.